\documentclass[aps,twocolumn,floatfix,showpacs,eqsecnum,prb]{revtex4}
\usepackage{graphicx}
\usepackage{amsmath}
\usepackage{amssymb}
\usepackage{bm}

\begin{document}

\title{Spin-triplet supercurrent carried by quantum Hall edge states through a Josephson junction}
\author{J. A. M. van Ostaay}
\affiliation{Instituut-Lorentz, Universiteit Leiden, P.O. Box 9506, 2300 RA Leiden, The Netherlands}
\author{A. R. Akhmerov}
\affiliation{Instituut-Lorentz, Universiteit Leiden, P.O. Box 9506, 2300 RA Leiden, The Netherlands}
\author{C. W. J. Beenakker}
\affiliation{Instituut-Lorentz, Universiteit Leiden, P.O. Box 9506, 2300 RA Leiden, The Netherlands}

\date{March, 2011}

\begin{abstract}
We show that a spin-polarized Landau level in a two-dimensional electron gas can carry a spin-triplet supercurrent between two spin-singlet superconductors. The supercurrent results from the interplay of Andreev reflection and Rashba spin-orbit coupling at the normal--superconductor (NS) interface. We contrast the current-phase relationship and the Fraunhofer oscillations of the spin-triplet and spin-singlet Josephson effect in the lowest Landau level, and find qualitative differences.
\end{abstract}

\maketitle

\section{Introduction}
\label{intro}

The coexistence of the quantum Hall effect with the superconducting proximity effect provides a unique opportunity to study the flow of supercurrent in chiral edge states. The usual quantum Hall edge states\cite{Hal82} in a two-dimensional (2D) electron gas are created by the interplay of cyclotron motion and reflection from an electrostatic potential, propagating in a direction dictated by the cyclotron frequency $\omega_{c}=eB/m$. At the interface with a superconductor Andreev reflection from the pair potential takes over, converting electrons into holes.\cite{And64} Since the sign of both the effective mass $m$ and charge $e$ change upon Andreev reflection, the cyclotron rotation keeps the same direction for electrons and holes and the chirality of these Andreev edge states is preserved.\cite{Tak98a,Hop00,Asa00a,Cht01}

While the superconducting proximity effect is short-ranged in the direction perpendicular to the edge states, it is long-ranged in the parallel direction. Indeed, a supercurrent can flow through a 2D electron gas even if the magnetic field is so strong that only a single Landau level is occupied --- provided the spin splitting by the Zeeman effect is sufficiently small.\cite{Ma93,Zyu94} Andreev reflection from a spin-singlet superconductor couples opposite spin bands, so spin polarization of the Landau level suppresses the supercurrent.\cite{Fis94,Moo99}

Recent studies of ferromagnetic Josephson junctions have shown that a spin-triplet proximity effect (with electrons and holes from the same spin band) can be induced by a spin-singlet superconductor, if the spin is not conserved at the ferromagnet-superconductor interface.\cite{Ber01,Kad01,Esc11} In the 2D electron gas of a quantum well formed in a narrow band gap semiconductor, such as InAs or InSb, the Rashba effect is a significant source of spin-orbit coupling in quantum Hall edge states.\cite{Pal05} When contacted with Nb electrodes, these structures show a strong proximity effect in the quantum Hall effect regime.\cite{Tak98b,Ero05,Bat07}

In this article we investigate whether the spin-polarized lowest Landau level of a 2D electron gas can carry a spin-triplet supercurrent between two spin-singlet superconductors, as a consequence of the Rashba effect on Andreev edge states. We find that a long-range spin-triplet proximity effect does exist, with a critical current $\propto(d/l_{\rm so})^{2}$, determined by the spin-orbit scattering length $l_{\rm so}$ in the normal region and the distance $d$ over which the electrostatic potential drops upon entering the superconductor. It is a small effect, but the fact that it exists as a matter of principle opens up the possibility to optimize it.

We calculate the current-phase relationship (dependence of the supercurrent on the superconducting phase difference) and the Fraunhofer oscillations (dependence on the magnetic flux through the junction) of the spin-triplet Josephson effect and compare with the corresponding spin-singlet effect. Some of our spin-singlet results are known\cite{Ma93,Zyu94,Sto11}, but some are new. In particular, we find a complete suppression of the Fraunhofer oscillations in the spin-singlet case for a critical value of the width $W$ of the Josephson junction. (These spin-singlet results may be of interest also for graphene, which shows a strong proximity effect\cite{Hee07} without significant spin-orbit coupling.) 

In Sec.\ \ref{spinpoltrans} we formulate the problem of edge state transport along a superconductor, in the form of an effective Hamiltonian in the lowest spin-split Landau level. The parameters entering into this Hamiltonian are derived from the Bogoliubov-De Gennes equation in the Appendix. The spin-triplet Josephson effect is analyzed in Secs.\ \ref{edgechJoseph} and \ref{spintriplet} and compared with the spin-singlet counterpart in Sec.\ \ref{spinsinglet}. We conclude in Sec.\ \ref{conclude}.

\section{Spin-polarized transport along a superconductor}
\label{spinpoltrans}

\subsection{NS interface}
\label{NSinterf}

\begin{figure}[tb]
\centerline{\includegraphics[width=0.8\linewidth]{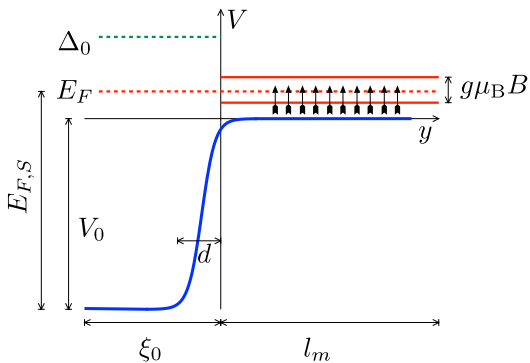}}
\caption{\label{fig_profile}
Schematic drawing of the energy scales and length scales at an NS interface. The electrostatic potential profile is shown as a blue solid curve, the Fermi level is a red dashed line, and the superconducting excitation gap is green dashed. The red solid lines indicate the spin-split lowest Landau level, with only a single spin band occupied (short black arrows).
}
\end{figure}

We consider the scattering by a superconductor (excitation gap $\Delta_{0}$, Fermi energy $E_{F,S}\gg \Delta_{0}$) of a single spin-polarized edge channel in a 2D electron gas in a perpendicular magnetic field $B$. The lowest Landau level at $\frac{1}{2}\hbar\omega_{c}\pm\frac{1}{2}g\mu_{\rm B}B$ is split by the Zeeman energy $g\mu_{\rm B}B$, and spin polarization is ensured by taking the Fermi level $E_{F}$ in the 2D gas in between the two spin-split levels (typically $E_{F}\approx\frac{1}{2}\hbar\omega_{c}$).

The characteristic energy and length scales at the normal--superconductor (NS) interface are shown in Fig.\ \ref{fig_profile}. On the superconducting side we have the coherence length $\xi_{0}=\hbar v_{F,S}/\Delta_{0}$, and the Fermi wave length $\lambda_{F,S}=2\pi/k_{F,S}=\pi\hbar v_{F,S}/E_{F,S}$. We require that $\xi_{0}$ is small compared to the magnetic length $l_{m}=\sqrt{\hbar/eB}$, to ensure that $B$ is well below the upper critical field of the superconductor.

The electrostatic potential step at the NS interface extends over a distance $d$, which we assume to be intermediate between $\lambda_{F,S}$ and $\xi_{0}$. These length scales are therefore ordered as
\begin{equation}
\lambda_{F,S}\ll d\ll \xi_{0}\ll l_{m}.\label{scalesordered}
\end{equation}
We include the rounding of the electrostatic potential step because it has a major effect on Andreev reflection. (For an abrupt interface, $d\lesssim\lambda_{F}$, Andreev reflection is strongly suppressed even without spin polarization.) The step in the pair potential is also rounded, but this has no significant effect on Andreev reflection (since $\Delta_{0}\ll E_{F,S}$).

On the normal side of the NS interface the Fermi wave length is $l_{m}$, so this is not an independent length scale. The spin-orbit scattering length and coupling energy are $l_{\rm so}=\hbar^{2}/m\alpha$ and $E_{\rm so}=m\alpha^{2}/\hbar^{2}$, respectively, with $\alpha$ the Rashba coefficient.

\subsection{Edge channel Hamiltonian}
\label{edgechannelH}

The wave function $\Psi=(\psi_{e},\psi_{h})$ of the electron and hole excitations (both in the same spin band) is an eigenstate of the Bogoliubov-De Gennes Hamiltonian $H$ with energy eigenvalue $\varepsilon$ (measured relative to the Fermi level). Electron-hole symmetry dictates that if $(\psi_{e},\psi_{h})$ is an eigenstate at energy $\varepsilon$ then $(\psi_{h}^{\ast},\psi_{e}^{\ast})$ is an eigenstate at energy $-\varepsilon$. This requires
\begin{equation}
\sigma_{x}H^{\ast}\sigma_{x}=-H,\label{ehsymm}
\end{equation}
where the Pauli matrix acts on the electron-hole degree of freedom.

At low excitation energies an effective Hamiltonian, containing only terms linear in momentum along the edge, is sufficient. The form of this effective Hamiltonian is fully constrained by the requirements of Hermiticity and electron-hole symmetry,
\begin{equation}
H=\frac{1}{2}\begin{pmatrix}
\{v_{c},p-eA\}&\{v_{\Delta},p\}\\
\{v_{\Delta}^{\ast},p\}&\{v_{c},p+eA\}
\end{pmatrix}.
\label{Hdef}
\end{equation}
Here $s$ and $\hat{s}$ are coordinate and unit vector along the edge, $p=-i\partial/\partial s$ is the canonical momentum, and $\bm{A}=A\hat{s}$ is the vector potential in a gauge where it is parallel to the edge. (We set $\hbar\equiv 1$ in intermediate formulas and write $+e$ for the electron charge.) The anti-commutator $\{a,b\}=ab+ba$ ensures that $H$ is Hermitian even if the velocities $v_{c}$ and $v_{\Delta}$ depend on $s$.

The gauge transformation $\Psi\mapsto \exp(i\chi\sigma_{z})\Psi$ transforms the Hamiltonian as follows,
\begin{align}
H\mapsto{}&e^{i\chi\sigma_{z}}He^{-i\chi\sigma_{z}}=\nonumber\\
&\frac{1}{2}\begin{pmatrix}
\{v_{c},p-eA-\chi'\}&\{|v_{\Delta}|e^{i\phi+2i\chi},p\}\\
\{|v_{\Delta}|e^{-i\phi-2i\chi},p\}&\{v_{c},p+eA+\chi'\}
\end{pmatrix},\label{gaugetrafo}
\end{align}
with $\chi'=\partial\chi/\partial s$ and $v_{\Delta}=|v_{\Delta}|e^{i\phi}$. We ensure that $v_{\Delta}$ is real positive by chosing $2\chi=-\phi$. The effective Hamiltonian then takes the form
\begin{equation}
H=(v_{c}+v_{\Delta}\sigma_{x})p-ev_{c}A\sigma_{z}-\tfrac{1}{2}i(v'_{c}+v'_{\Delta}\sigma_{x}).\label{HrealvDelta}
\end{equation}

\subsection{Dispersion relation}
\label{secdispersion}

\begin{figure}[tb]
\centerline{\includegraphics[width=0.8\linewidth]{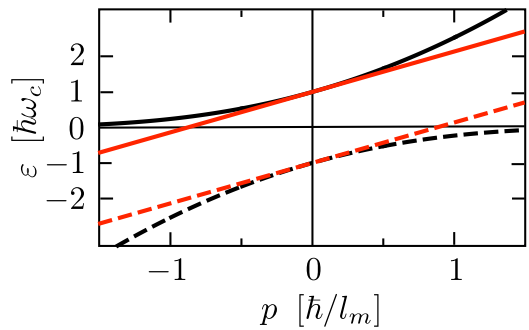}}
\caption{\label{fig_dispersion}
Dispersion relation of edge states along an NS interface in the lowest spin-polarized Landau level, for the electron-like mode (solid) and the hole-like mode (dashed). The black curves are calculated in the Appendix from the Bogoliubov-De Gennes equation (for $g\mu_{\rm B}B\ll\hbar\omega_{c}$, $v_{\Delta}/v_{c}\ll 1$, $\lambda/l_{m}\ll 1$). The red lines are the small-$p$ approximation \eqref{dispersion}.
}
\end{figure}

For $s$-independent $A$, $v_{c}$, and $v_{\Delta}$ the momentum $p$ along the edge is conserved. The Hamiltonian \eqref{HrealvDelta} describes two chiral modes with dispersion relation
\begin{equation}
\varepsilon=v_{c}p\pm\sqrt{(ev_{c}A)^{2}+(v_{\Delta}p)^{2}},\label{dispersion}
\end{equation}
see Fig.\ \ref{fig_dispersion}. At $\varepsilon=0$ the two modes have the same group velocity $v_{\rm group}=d\varepsilon/dp$, given for $v_{\Delta}\ll v_{c}$ by
\begin{equation}
v_{\rm group}=v_{c}-v_{\Delta}^{2}/v_{c}.
\end{equation}

Let us express $v_{c}$ and $v_{\Delta}$ in terms of the characteristic parameters of the NS interface. As derived in the Appendix, the two velocities $v_{c}$ and $v_{\Delta}$ are given, up to numerical coefficients of order unity, by
\begin{equation}
v_{c}\simeq l_{m}\omega_{c},\;\;v_{\Delta}\simeq \frac{v_{c}d}{l_{\rm so}}.\label{vcvDelta}
\end{equation}
The velocity $v_{c}$ is the same as the cyclotron drift velocity in the lowest Landau level along a normal, not superconducting boundary, in the limit of a steep confining potential. The confinement by the superconductor is effectively in that limit because the penetration depth $\xi_{0}$ of the edge state into the superconductor is less than its transverse extension $l_{m}$.

The velocity $v_{\Delta}$ which governs the coupling of electrons and holes is smaller than $v_{c}$ by a factor $d/l_{\rm so}$. Although it is the superconducting order parameter which scatters electrons into holes (Andreev reflection), the dependence on $\Delta_{0}$ drops out in the regime $\xi_{0}\ll l_{m}$. The ratio $d/l_{\rm so}$ appears in the calculation in the Appendix as the product of two factors, with a cancellation of the magnetic length: One factor is the probability $d/l_{m}$ of Andreev reflection with change of spin band and the other factor is the spin-flip probabiilty $l_{m}/l_{\rm so}$. The length $l_{\rm so}$ refers to spin-orbit scattering in N. There may also be spin-orbit scattering in S, but that would contribute to $v_{\Delta}$ a much smaller amount of order $v_{c}(d/l_{\rm so})(d/l_{m})^{2}$, see the Appendix.

\subsection{Effect of screening current}

The vector potential along the NS interface is determined by the screening of the magnetic field from the interior of the superconductor.\cite{Gia05} Consider an interface at $y=0$ with the superconductor in the region $y<0$. The edge state propagates in the $+x$ direction. The vector potential is $\bm{A}=A(y)\hat{x}$, with magnetic field $\bm{B}=-A'(y)\hat{z}$. We denote by $A_{0}=A(0)$ the value of $A$ at the NS interface. The Andreev-Rashba edge channel extends over a distance $l_{m}$ from the interface, so the effective value of the vector potential is $A_{\rm AR}= A_{0}-c_{m}l_{m}B$. The value of $c_{m}\approx 0.88$ is calculated in the Appendix.

The value of $A_{0}$ follows from the London equation for the screening supercurrent density $j$,
\begin{equation}
j=\frac{1}{2e\mu_{0}\lambda^{2}}\left(\frac{d\phi}{ds}-2eA_{0}\right),\label{Londoneq}
\end{equation}
with $\lambda$ the London penetration depth. For $l_{m}>\lambda$ the magnetic field decays exponentially $\propto e^{-|y|/\lambda}$ upon entering the superconductor. (This is the Meissner phase of a type-II superconductor, reached for magnetic fields below the lower critical field.) The screening current within a distance $\xi_{0}\ll\lambda$ from the interface is $j=B/\mu_{0}\lambda$. In the gauge where the order parameter is real, one thus has $A_{0}=-\lambda B$. 

We conclude that the vector potential $A$ in the edge state Hamiltonian \eqref{HrealvDelta} takes the value
\begin{equation}
A_{\rm AR}=-(c_{m}l_{m}+\lambda)B,\label{ABlambdalm}
\end{equation}
along the NS interface in the Meissner phase $l_{m}>\lambda$. The phase difference $2\pi |A_{\rm AR}|/\varphi_{0}$ accumulated per unit length between electron and hole (with $\varphi_{0}=h/2e$ the superconducting flux quantum) is increased by the screening current. This is a Doppler effect of Andreev reflection from a moving superconducting condensate.\cite{Gia05,Roh09}

For magnetic lengths in the range $\xi_{0}<l_{m}<\lambda$ the magnetic field penetrates into the superconductor in the form of Abrikosov vortices. In this mixed phase $A_{\rm AR}$ depends on the detailed configuration of vortices. We will consider this regime by treating $A_{\rm AR}$ as a random function of the position along the NS interface.

\subsection{Transfer matrix}
\label{transferM}

We transform from a Hamiltonian to a scattering description of the edge channel transport, which is the description we will use to calculate the Josephson current in a superconductor--normal-metal--superconductor (SNS) junction.

The particle current operator is
\begin{equation}
J=\partial H/\partial p=v_{c}+v_{\Delta}\sigma_{x}.\label{Jdef}
\end{equation}
We require $0\leq v_{\Delta}<v_{c}$, so $J^{1/2}$ is real Hermitian. To construct a unitary transfer matrix we transform $H$ to
\begin{equation}
\tilde{H}=J^{-1/2}HJ^{-1/2}=p-J^{-1/2}ev_{c}A\sigma_{z}J^{-1/2}.\label{tildeHdef}
\end{equation}
The wave function  $\tilde{\Psi}=J^{1/2}\Psi$ then satisfies
\begin{equation}
\tilde{H}\tilde{\Psi}=\varepsilon J^{-1}\tilde{\Psi}.\label{tildeHtildePsi}
\end{equation}

The transfer matrix $M(s_{2},s_{1})$ relates the function $\tilde{\Psi}(s)$ at two points along the boundary, $\tilde{\Psi}(s_{2})=M(s_{2},s_{1})\tilde{\Psi}(s_{1})$. Integration of Eq.\ \eqref{tildeHtildePsi} gives the expression
\begin{align}
&M(s_{2},s_{1})\nonumber\\
&\quad={\cal P}_{s}\exp\left[i\int_{s_{1}}^{s_{2}}ds
\left(\varepsilon J^{-1}+J^{-1/2}ev_{c}A\sigma_{z}J^{-1/2}\right)\right]\nonumber\\
&\quad={\cal P}_{s}\exp\left[i\int_{s_{1}}^{s_{2}}ds\left(\frac{\varepsilon(v_{c}-v_{\Delta}\sigma_{x})}{v_{c}^{2}-v_{\Delta}^{2}}+\frac{ev_{c}A\sigma_{z}}{\sqrt{v_{c}^{2}-v_{\Delta}^{2}}}\right)\right],
\label{Mdef}
\end{align}
with ${\cal P}_{s}$ the operator that orders the noncommuting matrices from left to right in order of decreasing $s$.

The transfer matrix is unitary, $M^{\dagger}=M^{-1}$, as an expression of particle current conservation: $\langle\tilde{\Psi}|\tilde{\Psi}\rangle=\langle\Psi|J|\Psi\rangle$ is independent of $s$. Electron-hole symmetry is expressed by
\begin{equation}
M|_{-\varepsilon}=\sigma_{x}M^{\ast}|_{\varepsilon}\,\sigma_{x}.\label{Mehsymmetry}
\end{equation}
 
Since the expression \eqref{Mdef} does not assume that the parameters $v_{c},v_{\Delta}$ are uniform along the edge, it may also be used to describe the transport along a boundary containing both normal and superconducting segments. On the normal segments $v_{\Delta}=0$ (no electron-hole coupling), while the cyclotron drift velocity $v_{c}$ is still given by Eq.\ \eqref{vcvDelta} (for a confining potential that is steep on the scale of $l_{m}$). The vector potential $A$ along the normal edge is determined by the enclosed magnetic flux, without the correction \eqref{ABlambdalm} from the screening current that is present along the superconducting edge.

Consider a superconducting segment connecting two normal boundaries. An electron enters the superconducting segment at the left end and exits at the right end, either as an electron or as a hole. At $\varepsilon=0$ the transfer matrix $M$ commutes with $\sigma_{z}$. This implies that, at the Fermi level, the electron exits as an electron with unit probability. At finite excitation energy $\varepsilon$ the probability for Andreev reflection (from electron to hole, with the transfer of a Cooper pair to the superconducting condensate) vanishes as $\varepsilon^{2}$ when $\varepsilon\rightarrow 0$, in accord with Refs.\ \onlinecite{Fis94,Ber09}.

\section{Edge channel Josephson effect}
\label{edgechJoseph}

\begin{figure}[tb]
\centerline{\includegraphics[width=0.8\linewidth]{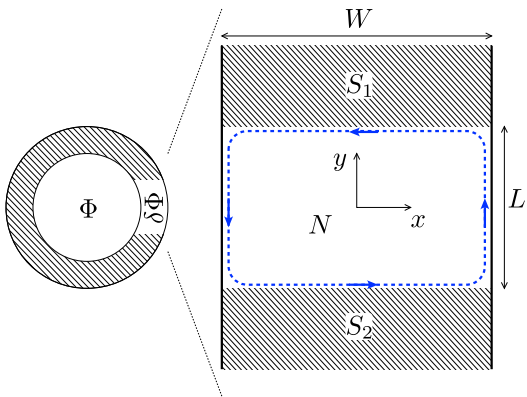}}
\caption{\label{fig_layout}
Left panel: Superconducting ring, enclosing a magnetic flux $\Phi$, interrupted in one arm by a normal segment (non-shaded region, containing a flux $\delta\Phi$). Right panel: enlargement of the SNS junction between the normal (N) and superconducting (S) regions. The normal region is a 2D electron gas in the quantum Hall effect regime, with a spin-polarized edge channel near the Fermi level (dashed, with arrows indicating the direction of motion).
}
\end{figure}

The geometry of the SNS Josephson junction is shown in Fig.\ \ref{fig_layout}. It consists of two parallel NS interfaces, interface $1$ at $y=L/2$ and interface $2$ at $y=-L/2$ (for both interfaces $|x|\leq W/2$). A wave incident on interface $1$ from point $s_{1}^{\rm in}=(W/2,0^{+})$ on the right edge comes out at point $s_{1}^{\rm out}=(-W/2,0^{+})$ on the left edge. The scattering matrix for this process is $S_{1}(\varepsilon)=M(s_{1}^{\rm out},s_{1}^{\rm in})|_{\varepsilon}$. Similarly, a wave incident on interface $2$ from point $s_{2}^{\rm in}=(-W/2,0^{-})$ on the left edge comes out at point $s_{2}^{\rm out}=(W/2,0^{-})$ on the right edge, with scattering matrix $S_{2}(\varepsilon)=M(s_{2}^{\rm out},s_{2}^{\rm in})|_{\varepsilon}$.

The SNS junction is a segment of a ring enclosing a magnetic flux $\Phi$, accounted for by a vector potential $\bm{A}_{\Phi}=\Phi\delta(y)\hat{y}$ (for $|x|\leq W/2$). The total scattering matrix $S(\varepsilon)$ for a closed scattering sequence, from $s_{1}^{\rm in}$ to $s_{1}^{\rm out}$ to $s_{2}^{\rm in}$ to $s_{2}^{\rm out}$ to $s_{1}^{\rm in}$, is given by
\begin{equation}
S(\varepsilon)=e^{i\pi\sigma_{z}\Phi/\varphi_{0}}S_{2}(\varepsilon)e^{-i\pi\sigma_{z}\Phi/\varphi_{0}}S_{1}(\varepsilon).\label{SS1S2}
\end{equation}

The contribution to the scattering matrix from the normal segments of the boundary can be calculated immediately from Eq.\ \eqref{Mdef}, because $v_{\Delta}=0$ and no operator ordering is required. We thus obtain
\begin{align}
S(\varepsilon)={}&e^{i\varepsilon\tau_{0}}e^{i(\pi/\varphi_{0})(\Phi+\delta\Phi/2)\sigma_{z}}M_{2}(\varepsilon)\nonumber\\
&\times e^{-i(\pi/\varphi_{0})(\Phi-\delta\Phi/2)\sigma_{z}}M_{1}(\varepsilon).\label{SS1S2lambda}
\end{align}
The flux through the junction is $\delta\Phi=BLW$ and $\tau_{0}=\oint ds\,v_{c}/(v_{c}^{2}-v_{\Delta}^{2})\approx 2(L+W)/v_{c}$ is the time it takes a quasiparticle to circulate along the entire perimeter of the junction. The matrices $M_{n}$ give the contribution to the scattering matrix from the interface with ${\rm S}_{n}$ (without the scalar factor, which has already been accounted for in the factor $e^{i\varepsilon\tau_{0}}$):
\begin{equation}
M_{n}(\varepsilon)={\cal P}_{s}\exp\left[i\int_{{\rm S}_{n}} ds\left(\frac{-\varepsilon v_{\Delta}\sigma_{x}}{v_{c}^{2}-v_{\Delta}^{2}}+\frac{ev_{c}A\sigma_{z}}{\sqrt{v_{c}^{2}-v_{\Delta}^{2}}}\right)\right].\label{Mndef}
\end{equation}

The Josephson current $I(\Phi)$ flowing in equilibrium at temperature $T$ through the SNS junction is related to the scattering matrix by\cite{Bro97}
\begin{equation}
I(\Phi)=\frac{1}{2} \frac{d}{d\Phi}\sum_{p=0}^{\infty}2k_{B}T\ln\,{\rm det}\,\bigl[1-S(i\omega_{p})\bigr]\label{IPhi}
\end{equation}
The imaginary energies are Matsubara frequencies, $i\omega_{p}=(2p+1)i\pi k_{B}T$. The prefactor $1/2$ accounts for the fact that only a single spin band contributes to the supercurrent. (In Ref.\ \onlinecite{Bro97} it is canceled by the spin degeneracy.) The Josephson current is a periodic function of the flux $\Phi$ through the ring, with period $\varphi_{0}$. The critical current $I_{c}$ of the Josephon junction is the largest value reached by $|I(\Phi)|$.

\section{Spin-triplet supercurrent}
\label{spintriplet}

\subsection{Calculation}
\label{sec_calculate}

To calculate the supercurrent we use the fact that $v_{\Delta}/v_{c}$ is a small parameter. An expansion in this parameter is made possible by the identity
\begin{align}
&{\cal P}_{s}\exp\left(\int_{0}^{W} ds\,[a(s)+b(s)]\right)=\nonumber\\
&\quad\quad A(W){\cal P}_{s}\exp\left(\int_{0}^{W}ds\,A^{-1}(s)b(s)A(s)\right),\label{operatoridentity}\\
& A(s)={\cal P}_{s'}\exp\left(\int_{0}^{s}ds'\,a(s')\right),\label{Asdef}
\end{align}
valid for any pair of operator functions $a(s)$, $b(s)$. An easy way to prove this identity is to call the right-hand-side $X(W)$ and calculate $dX/dW=[a(W)+b(W)]X(W)$. Integration then produces the left-hand-side. 

With the help of Eq.\ \eqref{operatoridentity}, the expression \eqref{Mndef} for the scattering matrix $M_{n}$ along the interface with ${\rm S}_{n}$ takes the form
\begin{align}
&M_{n}(\varepsilon)=e^{i\alpha_{n}\sigma_{z}}{\cal P}_{s}\exp\left[-i\varepsilon\int_{0}^{W} ds\,\frac{v_{\Delta}\sigma_{x}}{v_{c}^{2}-v_{\Delta}^{2}}e^{2iU_{n}\sigma_{z}}\right],\label{Mndef2}\\
&U_{n}(s)=\int_{0}^{s} ds'\,\frac{ev_{c}A_{n}(s')}{\sqrt{v_{c}^{2}-v_{\Delta}^{2}}},\;\;\alpha_{n}=U_{n}(W).\label{Unsdef}
\end{align}
The integral in the definition of the phase $U_{n}(s)$ runs over a distance $s$ along the ${\rm NS}_{n}$ interface, and $\alpha_{n}$ is the total phase accumulated by the vector potential $A_{n}(s)$ along that interface.

To first order in $v_{\Delta}$ the expression \eqref{Mndef2} reduces to
\begin{align}
M_{n}(\varepsilon)={}&e^{i\alpha_{n}\sigma_{z}}-i\varepsilon e^{i\alpha_{n}\sigma_{z}}\sigma_{x}\int_{0}^{W} ds\,(v_{\Delta}/v_{c}^{2})e^{2iU_{n}\sigma_{z}}\nonumber\\
={}&e^{i\alpha_{n}\sigma_{z}}\begin{pmatrix}
1&-i\varepsilon \Delta_{n}^{\ast}\\
-i\varepsilon \Delta_{n}&1
\end{pmatrix},\label{Mndef3}
\end{align}
with the definitions
\begin{align}
\Delta_{n}={}&\int_{0}^{W} ds\,\frac{v_{\Delta}}{v_{c}^{2}}\exp\left(2i\int_{0}^{s} ds'\,eA_{n}(s')\right),\label{Deltandef}\\
\alpha_{n}={}&\int_{0}^{W} ds\,eA_{n}(s).\label{alphansimple}
\end{align}

From Eq.\ \eqref{SS1S2lambda} we obtain the determinant, to second order in $v_{\Delta}$,
\begin{align}
&{\rm Det}\,[1-S(i\omega)]=2e^{-\omega\tau_{0}}\biggl[\cosh(\omega\tau_{0})\nonumber\\
&\quad-\cos(\pi\delta\Phi/\varphi_{0}+\alpha_{1}+\alpha_{2})-\tfrac{1}{2}e^{-\omega\tau_{0}}\omega^{2}\bigl(|\Delta_{1}|^{2}+|\Delta_{2}|^{2}\bigr)\nonumber\\
&\quad-\omega^{2}\,{\rm Re}\,\bigl(\Delta_{1}\Delta_{2}^{\ast}e^{i(\alpha_{2}-\alpha_{1}+2\pi\Phi/\varphi_{0})}\bigr)\biggr],\label{determinantrandom}
\end{align}
and then substitution into Eq.\ \eqref{IPhi} gives the supercurrent,
\begin{align}
&I(\Phi)=\frac{2\pi k_{B}T}{\varphi_{0}}{\rm Im}\,\bigl(\Delta_{1}\Delta_{2}^{\ast}e^{i(\alpha_{2}-\alpha_{1}+2\pi\Phi/\varphi_{0})}\bigr)\nonumber\\
&\quad\times\sum_{p=0}^{\infty}\omega_{p}^{2}\bigl[\cosh(\omega_{p}\tau_{0})-\cos(\pi\delta\Phi/\varphi_{0}+\alpha_{1}+\alpha_{2})\bigr]^{-1}.\label{IPhirandom}
\end{align}

This expression holds for arbitrary temperature and for arbitrary variation of $A(s)$, $v_{c}(s)$, and $v_{\Delta}(s)$ along the two NS interfaces, which is fully accounted for by the parameters $\Delta_{n}$ and $\alpha_{n}$ [Eqs.\ \eqref{Deltandef} and \eqref{alphansimple}]. We will now discuss this general result in some illustrative limits.

\subsection{High and low-temperature regimes}

The high-temperature limit ($k_{B}T\tau_{0}\gg 1$) of Eq.\ \eqref{IPhirandom} is given by the $p=0$ term in the sum over Matsubara frequencies,
\begin{equation}
I(\Phi)=4\pi^{2}e\sin(2\pi\Phi/\varphi_{0}+\psi)(k_{B}T)^{3}|\Delta_{1}\Delta_{2}|e^{-\pi k_{B}T\tau_{0}}.\label{IphihighT}
\end{equation}
The low-temperature limit is obtained by replacing the sum by an integration, with the result
\begin{align}
&I(\Phi)=\frac{e}{\pi}\frac{|\Delta_{1}\Delta_{2}|}{\tau_{0}^{3}}\sin(2\pi\Phi/\varphi_{0}+\psi){\cal F}(\pi\delta\Phi/\varphi_{0}+\psi'),\label{IphizeroT}\\
&\psi={\rm arg}\,\Delta_{1}-{\rm arg}\,\Delta_{2}+\alpha_{2}-\alpha_{1},\;\;\psi'=\alpha_{1}+\alpha_{2},\label{psidef}\\
&{\cal F}(x)=\int_{0}^{\infty}d\omega\,\frac{\omega^{2}}{\cosh\omega-\cos x}.\label{falphadef}
\end{align}
The function ${\cal F}(x)$ oscillates between ${\cal F}(0)=2\pi^{2}/3$ and ${\cal F}(\pi)=\pi^{2}/3$.

\begin{figure}[tb]
\centerline{\includegraphics[width=0.9\linewidth]{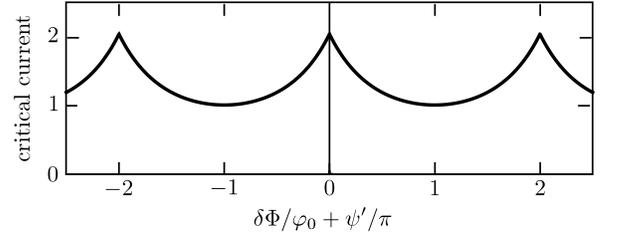}}
\caption{\label{fig_Icdphitriplet}
Low-temperature critical current $I_{c}$ as a function of the flux $\delta\Phi$ through the normal region, plotted from Eq.\ \eqref{IphizeroT} in units of $e|\Delta_{1}\Delta_{2}|/\tau_{0}^{3}$.
}
\end{figure}

The vector potential along the NS interfaces introduces a phase shift $\psi$ in the sinusoidal current-phase relationship, as a result of which the current $I(\Phi)$ is no longer and odd function of the flux $\Phi$ through the ring. 

In the high-temperature regime the critical current is $\delta\Phi$-independent, while at low temperatures it varies by a factor of two upon variation of $\delta\Phi$, see Fig.\ \ref{fig_Icdphitriplet}. The oscillations of the critical current as a function of the flux through the normal region are reminiscent of the Fraunhofer oscillations in a conventional Josephson junction,\cite{Tin96} but the minima are not at zero and the periodicity is $2\varphi_{0}$ rather than $\varphi_{0}$. These are characteristic signatures of a supercurrent carried by edge states rather than bulk states.\cite{Hei98,Bar99}

The low-temperature supercurrent decays $\propto 1/L^{3}$ if the separation $L$ of the NS interfaces is increased at constant width $W$. This is the characteristic decay of the spin-triplet proximity effect in a single transport channel.\cite{Ber09}

\subsection{Meissner phase}
\label{Meissner}

In the Meissner phase $l_{m}>\lambda$ we may take an $s$-independent vector potential $A_{\rm AR}$ along each NS interface, given by Eq.\ \eqref{ABlambdalm}. If we also take $s$-independent parameters $v_{c}$ and $v_{\Delta}$, the two quantities $\Delta_{n}$ and $\alpha_{n}$ defined in Eqs.\ \eqref{Deltandef} and \eqref{alphansimple} are given by
\begin{equation}
\Delta_{n}=\tau_{\Delta}e^{i\alpha_{n}}\frac{\sin\alpha_{n}}{\alpha_{n}},\;\;\alpha_{n}=\pi WA_{\rm AR}/\varphi_{0},\label{DeltaalphaMeissner}
\end{equation}
with $\tau_{\Delta}=Wv_{\Delta}/v_{c}^{2}$. (We kept the subscript $n$ to allow for possibly different values of $A_{\rm AR}$ at the two NS interfaces.)

The zero-temperature limit \eqref{IphizeroT} of the supercurrent then takes the form
\begin{align}
I(\Phi)={}&\frac{e}{\pi}\frac{\tau_{\Delta}^{2}}{\tau_{0}^{3}}\sin(2\pi\Phi/\varphi_{0}){\cal F}(\pi\delta\Phi/\varphi_{0}+\alpha_{1}+\alpha_{2})\nonumber\\
&\times\frac{\sin\alpha_{1}\sin\alpha_{2}}{\alpha_{1}\alpha_{2}},\label{IlambdazeroT}
\end{align}
with the function ${\cal F}$ defined in Eq.\ \eqref{falphadef}.

The phase shift in the $\Phi$-dependence has disappeared, so now the supercurrent is an odd function of the flux $\Phi$ through the ring, vanishing at $\Phi=0$. Since $dI/d\Phi>0$ at $\Phi=0$ (for $\alpha_{1}=\alpha_{2}$), the supercurrent is \textit{paramagnetic} --- in contrast with the usual diamagnetic Josephson effect. Such a $\pi$-junction appears generically in the spin-triplet proximity effect.\cite{Ber09} The main effect of the phase $\alpha_{n}$ accumulated by the vector potential along the NS interface is the reduction of the critical current by the factor $(\sin\alpha_{1}\sin\alpha_{2})/(\alpha_{1}\alpha_{2})$ --- the supercurrent vanishes if $\alpha_{1}$ or $\alpha_{2}$ is a (nonzero) integer multiple of $\pi$.

From Eq.\ \eqref{IlambdazeroT} we conclude that the scaling of the critical current with the parameters of the Josephson junction is given in the Meissner phase by
\begin{equation}
I_{c}\simeq \frac{e\tau_{\Delta}^{2}}{\tau_{0}^{3}}\frac{l_{m}^{2}}{W^{2}}\simeq e\omega_{c}(d/l_{\rm so})^{2}(l_{m}/{\cal L})^{3},\label{IcMeissner}
\end{equation}
with ${\cal L}=2(L+W)$ the length of the perimeter of the normal region. (All coefficients of order unity are disregarded in this scaling estimate, as well as any oscillatory dependence on $W$.)

\subsection{Mixed phase}
\label{mixed}

In the mixed phase  $\xi_{0}<l_{m}<\lambda$ the vector potential $A_{n}(s)$ along the NS interface depends on the configuration of vortices that have penetrated into the superconductor. There is now a random phase shift of the supercurrent, both as a function of $\Phi$ and $\delta\Phi$. The zero-temperature critical current reaches its maximal value $I_{c}^{\rm max}$ at $\delta\Phi=-(\alpha_{1}+\alpha_{2})\varphi_{0}/\pi$, given according to Eq.\ \eqref{IphizeroT} by
\begin{equation}
I_{c}^{\rm max}=\frac{2\pi e}{3\tau_{0}^{3}}|\Delta_{1}\Delta_{2}|.\label{Icmax}
\end{equation}
The $1/L^{3}$ scaling with the separation of the NS interfaces is unchanged, but the scaling with the width $W$ depends on the statistics of $A_{n}(s)$, which determines the statistics of $\Delta_{n}$ according to Eq.\ \eqref{Deltandef}.

We have calculated the average of $I_{c}^{\rm max}$ for a random variation of $A_{n}(s)$ as a function of $s$, with zero average and correlation length of order $l_{m}$ (the average separation of vortices). The magnitude of the fluctuations is quantified by taking a piecewise constant $A_{n}(s)$ in each segment of length $l_{m}$, drawn from a Gaussian distribution with zero average and standard deviation $\sigma\times\varphi_{0}/\pi l_{m}$ with $\sigma$ of order unity. We have found that the average critical current in the mixed phase scales for $W\gg l_{m}$ as
\begin{equation}
\bigl\langle I_{\rm c}^{\rm max}\bigr\rangle\simeq \frac{e\tau_{\Delta}^{2}}{\tau_{0}^{3}}\frac{l_{m}}{W}\simeq e\omega_{c}\frac{d^{2}}{l_{\rm so}^{2}}\frac{l_{m}^{2}W}{{\cal L}^{3}},\label{Icmaxaverage}
\end{equation}
larger than in the Meissner phase by a factor $W/l_{m}$.

\section{Comparison with spin-singlet supercurrent}
\label{spinsinglet}

\subsection{Transfer matrix}
\label{sec_transfersinglet}

It is instructive to compare the results of the previous section for the spin-triplet supercurrent with the spin-singlet case considered by Ma and Zyuzin.\cite{Ma93,Zyu94} For that purpose we assume spin degeneracy in the 2D electron gas, neglecting Zeeman splitting or spin-orbit coupling. Electron-hole symmetry now relates excitations from opposite spin bands, say an electron from the spin-up band and a hole from the spin-down band (or vice versa). 

The effective Hamiltonian of a spin-singlet edge channel, to linear order in momentum, is
\begin{equation}
H=\begin{pmatrix}
\frac{1}{2}\{v_{c},p-eA\}&\Delta\\
\Delta^{\ast}&\frac{1}{2}\{v_{c},p+eA\}
\end{pmatrix},
\label{Hdefsinglet}
\end{equation}
fully constrained by Hermiticity and the electron-hole symmetry requirement
\begin{equation}
\sigma_{y}H^{\ast}\sigma_{y}=-H.\label{ehsymmsinglet}
\end{equation}
Choosing a gauge so that $\Delta$ is real we now have
\begin{equation}
H=v_{c}(p-eA\sigma_{z})+\Delta\sigma_{x}-\tfrac{1}{2}iv'_{c}.\label{HrealvDeltasinglet}
\end{equation}
The key difference with the effective Hamiltonian \eqref{HrealvDelta} for a spin-triplet edge channel is that the coupling between electrons and holes does not vanish at $p=0$ in the spin-singlet case.

We now follow the same steps as in Sec.\ \ref{transferM}. The particle current operator
\begin{equation}
J=\partial H/\partial p=v_{c}\label{Jdefsinglet}
\end{equation}
transforms $H$ to
\begin{equation}
\tilde{H}=J^{-1/2}HJ^{-1/2}=p-eA\sigma_{z}+(\Delta/v_{c})\sigma_{x},\label{tildeHdefsinglet}
\end{equation}
and produces the unitary transfer matrix
\begin{equation}
M(s_{2},s_{1})={\cal P}_{s}\exp\left[i\int_{s_{1}}^{s_{2}}ds\left(\frac{\varepsilon-\Delta\sigma_{x}}{v_{c}}+eA\sigma_{z}\right)\right].
\label{Mdefsinglet}
\end{equation}
The transfer matrix no longer commutes with $\sigma_{z}$ at $\varepsilon=0$, so there is no low-energy suppression of Andreev reflection as in the spin-triplet case. The order parameter $\Delta$ equals $\Delta_{0}$ along the NS interface and zero along the normal boundary.

\subsection{Meissner phase}
\label{Meissner_singlet}

We consider the Meissner phase $l_{m}>\lambda$, with an $s$-independent vector potential $A_{n}$ along the interface with ${\rm S}_{n}$. Taking also an $s$-independent $v_{c}$, we can evaluate Eq.\ \eqref{Mdefsinglet} without the complications from operator ordering. The scattering matrix becomes
\begin{align}
S(\varepsilon)={}&e^{i\varepsilon\tau_{0}}e^{i(\pi/\varphi_{0})(\Phi+\delta\Phi/2)\sigma_{z}}\tilde{M}_{2}\nonumber\\
&\times e^{-i(\pi/\varphi_{0})(\Phi-\delta\Phi/2)\sigma_{z}}\tilde{M}_{1},\label{SS1S2lambdasinglet}\\
\tilde{M}_{n}={}&\exp\bigl[ieWA_{n}\sigma_{z}-i(\Delta_{0}W/v_{c})\sigma_{x}\bigr],\label{Mnsinglet}
\end{align}
with $\tau_{0}=2(L+W)/v_{c}$. The supercurrent follows from
\begin{equation}
I(\Phi)= \frac{d}{d\Phi}\sum_{p=0}^{\infty}2k_{B}T\ln\,{\rm det}\,\bigl[1-S(i\omega_{p})\bigr],\label{IPhi2}
\end{equation}
which differs from Eq.\ \eqref{IPhi2} by a factor of $2$ because of spin degeneracy of the edge channel in the spin-singlet case.

Substitution of Eq.\ \eqref{SS1S2lambdasinglet} into Eq.\ \eqref{IPhi2} gives
\begin{align}
I(\Phi)={}&-\frac{4\pi k_{B}T}{\varphi_{0}}\sin(2\pi\Phi/\varphi_{0})(W/\xi_{c})^{2}\frac{\sin^{2}\beta}{\beta^{2}}\nonumber\\
&\times\sum_{p=0}^{\infty}\bigl[\cosh(\omega_{p}\tau_{0})+X\bigr]^{-1},\label{Iphiresultsinglet}\\
X={}&[\cos(2\pi\Phi/\varphi_{0})-\cos(\pi\delta\Phi/\varphi_{0})](W/\xi_{c})^{2}\frac{\sin^{2}\beta}{\beta^{2}}\nonumber\\
&+(\pi WA_{\rm AR}/\varphi_{0})\frac{\sin 2\beta}{\beta}\sin(\pi\delta\Phi/\varphi_{0})\nonumber\\
&-\cos 2\beta\cos(\pi\delta\Phi/\varphi_{0}),\label{Xdef}\\
\beta={}&\sqrt{(\pi WA_{\rm AR}/\varphi_{0})^{2}+(W/\xi_{c})^{2}}.\label{betadef}
\end{align}
(For a compact expression, we took $A_{1}=A_{2}\equiv A_{\rm AR}$.) We defined the length $\xi_{c}=\hbar v_{c}/\Delta_{0}$, smaller than the superconducting coherence length $\xi_{0}=\hbar v_{F,S}/\Delta_{0}$ by a factor $v_{c}/v_{F,S}$. In the point contact limit $W\rightarrow 0$ considered by Ma and Zyuzin our result \eqref{Iphiresultsinglet} agrees with their finding (Eq.\ 13 of Ref.\ \onlinecite{Zyu94}).

At zero temperature Eq.\ \eqref{Iphiresultsinglet} evaluates to
\begin{align}
I(\Phi)={}&-\frac{4e}{\pi\tau_{0}}\sin(2\pi\Phi/\varphi_{0})(W/\xi_{c})^{2}\frac{\sin^{2}\beta}{\beta^{2}}\nonumber\\
&\times\frac{1}{\sqrt{1-X^{2}}}\,{\rm arctan}\,\left(\frac{1-X}{\sqrt{1-X^{2}}}\right).\label{IPhisingletT0}
\end{align}
In contrast to the spin-singlet result \eqref{IphizeroT}, the dependence of the supercurrent on the flux $\Phi$ through the ring is strongly non-sinusoidal. The critical current oscillates both as a function of the flux $\delta\Phi$ through the normal region and as a function of the width $W$ of the NS interface. At high temperature only the oscillation with $W$ remains,
\begin{equation}
I(\Phi)=-8e k_{B}T\sin(2\pi\Phi/\varphi_{0})(W/\xi_{c})^{2}\frac{\sin^{2}\beta}{\beta^{2}}e^{-\pi k_{B}T\tau_{0}},\label{IPhisinglethighT}
\end{equation}
while the $\Phi$-dependence is now sinusoidal.

Upon increasing the separation $L$ of the NS interfaces the spin-singlet supercurrent \eqref{IPhisingletT0} in the low-temperature limit decays as $1/L$. This is in contrast to the $1/L^{3}$ decay of the spin-triplet supercurrent \eqref{IphizeroT}. In the high-temperature limit the supercurrent has the same exponential decay $\propto\exp(-\pi k_{B}T\tau_{0})$ in the spin-singlet and spin-triplet cases --- only the pre-exponentials differ [cf.\ Eqs.\ \eqref{IphihighT} and \eqref{IPhisinglethighT}].

The spin-singlet supercurrent in the high-temperature regime has been studied also by Ishikawa and Fukuyama,\cite{Ish99} without taking the point contact limit $W\rightarrow 0$ of Refs.\ \onlinecite{Ma93,Zyu94}. We have not been able to reconcile their result with our Eq.\ \eqref{IPhisinglethighT}, because only the length $L$ of the normal boundaries enters into their exponential decay (rather than the sum $L+W$ of the lengths of normal and superconducting boundaries). The very recent study by Stone and Lin,\cite{Sto11} which also includes finite-$W$ effects, still assumes $W\ll L$ so it does not distinguish between the two decay rates.

\subsection{Narrow-contact regime}
\label{narrowcontact}

\begin{figure}[tb]
\centerline{\includegraphics[width=0.9\linewidth]{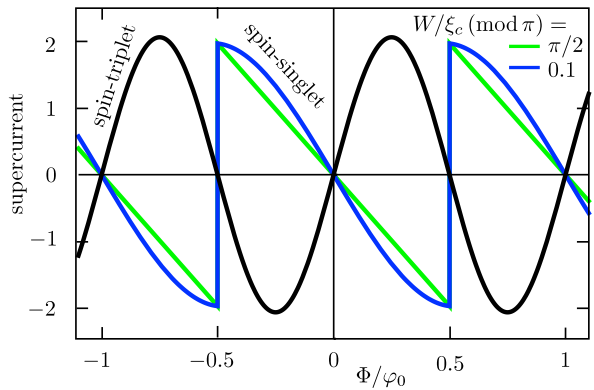}}
\caption{\label{fig_IPhi}
Zero-temperature supercurrent as a function of the flux $\Phi$ through the ring, for a flux $\delta\Phi$ through the normal region equal to an integer multiple of $2\varphi_{0}=h/e$. The green and blue curves (in units of $(e/\tau_{0})|\sin W/\xi_{c}|$) are the spin-singlet result \eqref{IPhisingletT0} for two values of $W$ (in the narrow-contact regime $W\ll l_{m}$, so with $A_{\rm AR}\rightarrow 0$). The black curve is the spin-triplet result \eqref{IphizeroT}, plotted in units of $e\tau_{\Delta}^{2}/\tau_{0}^{3}$. For the sake of comparison we also took the narrow-contact limit of the spin-triplet result, setting $\alpha_{1},\alpha_{2}\rightarrow 0$ in Eq.\ \eqref{IphizeroT}.
}
\end{figure}

\begin{figure}[tb]
\centerline{\includegraphics[width=0.9\linewidth]{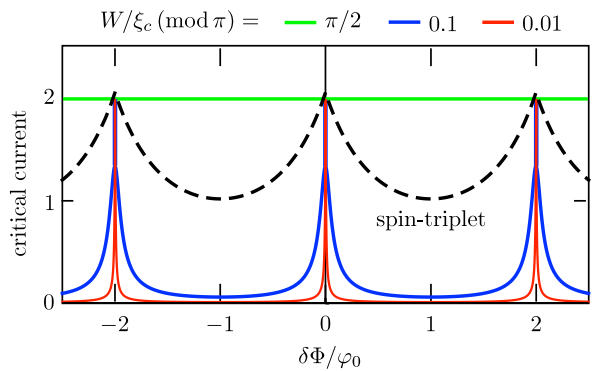}}
\caption{\label{fig_Icdphi}
Low-temperature critical current as a function of the flux $\delta\Phi$ through the normal region. The dashed curve is the spin-triplet result \eqref{IphizeroT}, plotted in units of $e\tau_{\Delta}^{2}/\tau_{0}^{3}$ (in the narrow-contact limit $\alpha_{1},\alpha_{2}\rightarrow 0$). The solid curves (in units of $(e/\tau_{0})|\sin W/\xi_{c}|$) follow from the spin-singlet result \eqref{IPhisingletT0} for three values of $W$ in the narrow-contact regime. The resonance at integer $\delta\Phi/2\varphi_{0}$ peaks at $I_{c}=(2\pi/3)e\tau_{\Delta}^{2}/\tau_{0}^{3}$ in the spin-triplet case and at $I_{c}=(2e/\tau_{0})|\sin W/\xi_{c}|$ in the spin-singlet case.
}
\end{figure}

The full expression \eqref{IPhisingletT0} for the zero-temperature spin-singlet supercurrent simplifies considerably in the narrow-contact regime $W\ll l_{m}$, when we may set $A_{\rm AR}\rightarrow 0$. (This is the regime considered by Stone and Lin.\cite{Sto11}) Note that $\xi_{c}/l_{m}\simeq \hbar\omega_{c}/\Delta_{0}\ll 1$, so $W$ may still be large compared to $\xi_{c}$ in the narrow-contact regime. As shown in Fig.\ \ref{fig_IPhi}, the current-phase relationship in the narrow-contact regime has a sawtooth-like shape, consistent with Ref.\ \onlinecite{Sto11}.

For reduced width $w\equiv W/\xi_{c}$ (modulo $\pi$) much less than unity the critical current exhibits resonant peaks (of height $2ew/\tau_{0}$) whenever $\delta\Phi/2\varphi_{0}$ is an integer. (See Fig.\ \ref{fig_Icdphi}, blue and red curves.) For $w\rightarrow\pi/2$ the critical current $I_{c}=2e/\tau_{0}$ becomes $\delta\Phi$-independent (green line in Fig.\ \ref{fig_Icdphi}) --- signifying the absence of Fraunhofer oscillations.

\section{Conclusion}
\label{conclude}

In conclusion, we have analysed the Josephson effect in the lowest Landau level, both with and without spin-polarization. The critical current scales differently with the parameters of the Josephson junction in these two cases. Without spin-polarization we have the spin-singlet Josephson effect considered earlier,\cite{Ma93,Zyu94,Sto11} with low-temperature scaling
\begin{equation}
I_{\rm c,singlet}\simeq e\omega_{c}\frac{l_{m}}{{\cal L}},\label{Icsinglet}
\end{equation}
inversely proportional to the length ${\cal L}$ of the perimeter of the normal region.

We have found that a spin-polarized Landau level can still carry a supercurrent. The low-temperature scaling of this spin-triplet Josephson effect is
\begin{equation}
I_{\rm c,triplet}\simeq e\omega_{c}\left(\frac{l_{m}}{\cal L}\right)^{3}(W/l_{m})(d/l_{\rm so})^{2},\label{Ictriplet}
\end{equation}
in the mixed phase with $W\gg l_{m}$.

For $W\simeq {\cal L}$ the ratio of spin-triplet and spin-singlet critical currents is of order
\begin{equation}
I_{\rm c,triplet}/I_{\rm c,singlet}\simeq (l_{m}/{\cal L})(d/l_{\rm so})^{2}.\label{ratiosinglettriplet}
\end{equation}
The spin-orbit scattering length in InAs is of order $l_{\rm so}\simeq 100\,{\rm nm}$, which could well be of the same order as the electrostatic length $d$ (the smoothness of the potential step at the NS interface). The main reason for the relative smallness of the spin-triplet supercurrent is then the factor $l_{m}/{\cal L}$. Since $l_{m}\lesssim 25\,{\rm nm}$ for $B\gtrsim 1\,{\rm T}$, a submicron junction is needed for an observable effect. As we have shown, the spin-triplet Josephson effect has unusual features, including a paramagnetic, rather than diamagnetic, current-phase relationship, and Fraunhofer oscillations which have a $h/e$ rather than $h/2e$ periodicity.

For the purpose of comparison with the spin-singlet Josephson effect, we have performed an analysis that goes beyond earlier work on that problem,\cite{Ma93,Zyu94,Sto11} in particular with regard to the Fraunhofer oscillations. We have found a remarkable dependence of the amplitude of the Fraunhofer oscillations on the relative magnitude of the junction width $W$ and an effective coherence length $\xi_{c}$. For $W/\xi_{c}=\pi/2$ (mod $\pi$) the Fraunhofer oscillations vanish alltogether, see Fig.\ \ref{fig_Icdphi}. 

These spin-singlet results may well be of relevance also for graphene, which is an attractive alternative to InAs in the search for the coexistence of the Josephson and quantum Hall effects. The results obtained here would apply if $W$ is larger than the intervalley scattering length. For smaller $W$ the valley-selectivity of the edge states enters, along the lines described in Ref.\ \onlinecite{Akh07}.

\acknowledgments

This research was supported by the Dutch Science Foundation NWO/FOM, by the Eurocores program EuroGraphene, and by an ERC Advanced Investigator grant.

\appendix

\section{Andreev-Rashba edge states}
\label{AndreevRashba}

The theory of Andreev edge states, produced by the interplay of cyclotron motion and Andreev reflection, has been developed by Z\"{u}licke and collaborators.\cite{Gia05,Hop00,Zul01} Here we include the interplay with Rashba spin-orbit interaction, in the spin-polarized regime where Andreev reflection can only occur because of the Rashba effect. 

The theory is complicated by the fact that we are deep in the quantum mechanical regime, with only one occupied Landau level, and cannot make the semiclassical approximation of large Landau level index made in earlier work.\cite{Gia05,Hop00,Zul01,Rak07} Since the Fermi energy in the normal metal is small compared to the superconducting gap, we can also not make the usual Andreev approximation (matching wave amplitudes without matching derivatives). We keep the theory tractable analytically by treating the spin-orbit interaction perturbatively.

The goal of our analysis of the Andreev-Rashba edge states is to arrive at a microscopic derivation of the parameters that enter into the effective edge state Hamiltonian \eqref{Hdef}, on which our theory of the spin-triplet Josephson effect is based.

\subsection{Bogoliubov-De Gennes equation}
\label{sec_BdGeq}

We start from the Bogoliubov-De Gennes (BdG) equation
\begin{equation}
\begin{pmatrix}
H_{0}-E_{F}&\Delta\tau_{y}\\
\Delta^{\ast}\tau_{y}&E_{F}-H_{0}^{\ast}
\end{pmatrix}\begin{pmatrix}
\psi_{e}\\
\psi_{h}
\end{pmatrix}=
\varepsilon\begin{pmatrix}
\psi_{e}\\
\psi_{h}
\end{pmatrix},\label{BdGeq}
\end{equation}
for quasiparticle excitations consisting of an electron spinor $\psi_{e}=(u_{+},u_{-})$ and a hole spinor $\psi_{h}=(v_{+},v_{-})$. The label $\pm$ indicates the spin band and the Pauli matrix $\tau_{y}$ acts on the spin degree of freedom. The pair potential $\Delta$ of a spin-singlet superconductor couples electron and hole excitations in opposite spin bands. Electron-hole symmetry is expressed by $\sigma_{x}H^{\ast}\sigma_{x}=-H$.

The single-particle Hamiltonian
\begin{align}
&H_{0}=(2m)^{-1}(\bm{p}-e\bm{A})^{2}+V+\tfrac{1}{2}g\mu_{\rm B}\bm{B}\cdot\bm{\tau}+H_{\rm R},\label{H0HR}\\
&H_{\rm R}=\alpha(p_{y}-eA_{y})\tau_{x}-\alpha(p_{x}-eA_{x})\tau_{y},\label{HRdef}
\end{align}
contains the kinetic energy, potential energy, Zeeman energy, and Rashba spin-orbit interaction. We consider a translationally invariant NS interface at $y=0$, with vector potential $\bm{A}=A(y)\hat{x}$, magnetic field $\bm{B}=-A'(y)\hat{z}$, electrostatic potential $V=V(y)$, and pair potential $\Delta=\Delta(y)$. The effective mass $m$, effective gyromagnetic factor $g$, and Rashba coefficient $\alpha$ are taken spatially uniform (otherwise also derivatives of $m$ and $\alpha$ would have to enter in the Hamiltonian, to preserve Hermiticity). 

Parallel momentum $p_{x}\equiv p$ is conserved, for states $\propto e^{ipx}$. The $y$-dependence of the wave functions is determined by
\begin{align}
H_{0}=&-\frac{1}{2m}\frac{d^{2}}{dy^{2}}+\frac{[p-eA(y)]^{2}}{2m}+V(y)\nonumber\\
&-\tfrac{1}{2}g\mu_{\rm B}A'(y)\tau_{z}+H_{\rm R},\label{H0RHp}\\
H_{\rm R}=&-i\alpha\tau_{x}\frac{d}{dy}-\alpha [p-eA(y)]\tau_{y}.\label{HRp}
\end{align}
In this basis the operators $H_{0},H_{0}^{\ast}$ in the BdG Hamiltonian should be replaced by $H_{0}(p)$, $H_{0}^{\ast}(-p)$.

The NS interface is at $y=0$, with the superconductor in the region $y<0$. In the simplest model for the interface we take a step function both for the pair potential, $\Delta(y)=\Delta_{0}\theta(-y)$, and for the electrostatic potential, $V(y)=-V_{0}\theta(-y)$ with $V_{0}>0$. (The function $\theta(y)$ equals $1$ for $y>0$ and $0$ for $y<0$.) Smoothing of the interface is important, and will be considered at the end of the Appendix. We assume that we are deep in the Meissner phase, $l_{m}\gg\lambda$, so that we may neglect the penetration of the magnetic field in the superconductor. In the gauge where $\Delta_{0}$ is real, the vector potential is then given simply by $A(y)=-yB\theta(y)$.
 
We will first solve the eigenvalue problem to zeroth order for $H_{\rm R}=0$, and then include the Rashba spin-orbit interaction to lowest order as a perturbation.

\subsection{Solution without the Rashba effect}
\label{sec_zero}

\subsubsection{Eigenstates in S}
\label{eigenS}

In S (for $y<0$) the BdG Hamiltonian with $H_{\rm R}=0$ is given by
\begin{widetext}
\begin{equation}
H_{S}=
\begin{pmatrix}
\mu(p)-\kappa\partial^{2}_{y}&0&0&-i\Delta_{0}\\
0&\mu(p)-\kappa\partial^{2}_{y}&i\Delta_{0}&0\\
0&-i\Delta_{0}&-\mu(p)+\kappa\partial^{2}_{y}&0\\
i\Delta_{0}&0&0&-\mu(p)+\kappa\partial^{2}_{y}
\end{pmatrix},\label{HS0}
\end{equation}
\end{widetext}
with $\mu(p)=p^{2}/2m-V_{0}-E_{F}$, $\kappa=(2m)^{-1}$, and $\partial_{y}=d/dy$.

There are four eigenstates $\chi_{s,\pm}w_{\pm}(y)$ of $H_{S}$ for $0<\varepsilon<\Delta_{0}$ (decaying for $y\rightarrow-\infty$), with
\begin{equation}
w_{\pm}(y)=e^{iq_{\pm}y},\;\;\gamma_{\pm}=\varepsilon\pm
i\sqrt{\Delta_{0}^{2}-\varepsilon^{2}},\label{wdef}
\end{equation}
\begin{align}
&\kappa q_{\pm}^{2}=-\mu(p)\pm i\sqrt{\Delta_{0}^{2}-\varepsilon^{2}},\;\;{\rm
Im}\,q_{\pm}<0,\label{qpmdef}\\
&\chi_{\uparrow,\pm}=\begin{pmatrix}
\gamma_{\pm}\\
0\\
0\\
i\Delta_{0}
\end{pmatrix},\;\;
\chi_{\downarrow,\pm}=\begin{pmatrix}
0\\
i\Delta_{0}\\
\gamma_{\mp}\\
0
\end{pmatrix}.\label{chipmdef}
\end{align}
For $\Delta_{0}\ll E_{F}+V_{0}\equiv E_{F,S}\equiv p_{F,S}^{2}/2m$ we may
approximate
\begin{equation}
q_{\pm}=\mp q(p)-\frac{im}{q(p)}\sqrt{\Delta_{0}^{2}-\varepsilon^{2}},\;\; q(p)=\sqrt{p_{F,S}^{2}-p^{2}}.\label{qpmsimple}
\end{equation}

\subsubsection{Eigenstates in N}
\label{eigenN}

In N (for $y>0$) we have, again for $H_{\rm R}=0$,
\begin{widetext}
\begin{equation}
H_{N}=
\begin{pmatrix}
U(p,y)-\kappa\partial^{2}_{y}+\mu_{+}&0&0&0\\
0&U(p,y)-\kappa\partial^{2}_{y}+\mu_{-}&0&0\\
0&0&-U(-p,y)+\kappa\partial^{2}_{y}-\mu_{+}&0\\
0&0&0&-U(-p,y)+\kappa\partial^{2}_{y}-\mu_{-}
\end{pmatrix},\label{HN0}
\end{equation}
with $\mu_{\pm}=\pm\frac{1}{2}g\mu_{\rm B}B-E_{F}$ and $U(p,y)=(p+eBy)^{2}/2m$.
\end{widetext}

The differential equation
\begin{equation}
[U(p,y)-\kappa\partial^{2}_{y}+\mu_{\pm}]\phi(y)=\varepsilon\phi(y),\label{Uphipm}
\end{equation}
with $\phi(y)\rightarrow 0$ for $y\rightarrow-\infty$ is solved by a parabolic cylinder function ${\cal U}$, 
\begin{equation}
\phi_{\pm}(\varepsilon,p,y)=C^{\pm}_{\varepsilon,p}\,{\cal U}\left[\frac{\mu_{\pm}-\varepsilon}{\omega_{c}},\sqrt{2}\left(\frac{y}{l_{m}}+pl_{m}\right)\right].\label{phipmcalU}
\end{equation}
The normalization constant $C^{\pm}_{\varepsilon,p}={\cal O}(l_{m}^{-1/2})$ is determined by
\begin{equation}
\int_{0}^{\infty}\phi_{\pm}^{2}(\varepsilon,p,y)\,dy=1.\label{phipmnormalize}
\end{equation}
The parabolic cylinder function ${\cal U}(-\nu,y)$ has no nodes as a function of $y$ for $\nu\leq 1/2$ and only a single node for $1/2<\nu\leq 3/2$.

The four eigenstates of $H_{N}$ are constructed in terms of the functions $\phi_{\pm}$,
\begin{align}
&\psi_{e\uparrow}=\begin{pmatrix}
\phi_{+}(\varepsilon,p,y)\\
0\\
0\\
0
\end{pmatrix},\;\;
\psi_{e\downarrow}=\begin{pmatrix}
0\\
\phi_{-}(\varepsilon,p,y)\\
0\\
0
\end{pmatrix},\nonumber\\
&\psi_{h\uparrow}=\begin{pmatrix}
0\\
0\\
\phi_{+}(-\varepsilon,-p,y)\\
0
\end{pmatrix},\;\;
\psi_{h\downarrow}=\begin{pmatrix}
0\\
0\\
0\\
\phi_{-}(-\varepsilon,-p,y)
\end{pmatrix}.\label{psiehdef}
\end{align}

\subsubsection{Matching at the NS interface}
\label{NSmatch}

We construct two independent superpositions of basis states in N and S,
\begin{subequations}
\label{Psi12def}
\begin{align}
\Psi_{1}(y)={}&\theta(y)\left[a_{1}\psi_{h,\uparrow}(y)+b_{1}\psi_{e,\downarrow}
(y)\right]\nonumber\\
&+\theta(-y)\left[c_{1}\chi_{\downarrow,+}e^{iq_{+}y}+d_{1}\chi_{\downarrow,-}e^{iq_{-}y}\right],\label{Psi2def}\\
\Psi_{2}(y)={}&\theta(y)\left[a_{2}\psi_{e,\uparrow}(y)+b_{2}\psi_{h,\downarrow}
(y)\right]\nonumber\\
&+\theta(-y)\left[c_{2}\chi_{\uparrow,+}e^{iq_{+}y}+d_{2}\chi_{\uparrow,-}e^{iq_{-}y}\right].\label{Psi1def}
\end{align}
\end{subequations}

We choose $\varepsilon$ such that
\begin{equation}
(\varepsilon-\mu_{+})/\hbar\omega_{c}<\tfrac{1}{2}<(\varepsilon-\mu_{-})/\hbar\omega_{c}<\tfrac{3}{2}.\label{range}
\end{equation}
In this range the equation $\phi_{+}(\varepsilon,p,0)=0$ has no solution while the equation $\phi_{-}(\varepsilon,p,0)=0$ has a single solution $p={\cal D}(\varepsilon)$. As we will see, this is the branch of the dispersion relation with wave function $\Psi_{1}$, while another branch, with wave function $\Psi_{2}$, is given by $p=-{\cal D}(-\varepsilon)$.

Continuity of $\Psi_{1}$ and $d\Psi_{1}/dy$ at $y=0$ gives four equations for the coefficients $a_{1},b_{1},c_{1},d_{1}$, 
\begin{subequations}
\label{phi2eq}
\begin{align}
&a_{1}\phi_{+}(-\varepsilon,-p,0)=c_{1}\gamma_{-}+d_{1}\gamma_{+},\label{phi2a}\\
&b_{1}\phi_{-}(\varepsilon,p,0)=i\Delta_{0}(c_{1}+d_{1}),\label{phi2b}\\
&a_{1}\phi'_{+}(-\varepsilon,-p,0)=iq_{+}c_{1}\gamma_{-}+iq_{-}d_{1}\gamma_{+},\label{phi2a2}\\
&b_{1}\phi'_{-}(\varepsilon,p,0)=-\Delta_{0}(q_{+}c_{1}+q_{-}d_{1}),\label{phi2b2}
\end{align}
\end{subequations}
with $\phi'_{\pm}=d\phi_{\pm}/dy$. The solution satisfies
\begin{align}
\frac{c_{1}}{d_{1}}&=-\frac{\gamma_{+}}{\gamma_{-}}\,\frac{q_{-}\phi_{+}(-\varepsilon,-p,0)+i\phi'_{+}(-\varepsilon,-p,0)}{q_{+}\phi_{+}(-\varepsilon,-p,0)+i\phi'_{+}(-\varepsilon,-p,0)}\nonumber\\
&=-\frac{\gamma_{+}q_{-}}{\gamma_{-}q_{+}}\left[1+{\cal O}(\lambda_{F,S}/l_{m})\right],\label{solutionsatisfies2}
\end{align}
since $\phi'_{+}$ is smaller than $q_{\pm}\phi_{+}$ by a factor $\lambda_{F,S}/l_{m}\ll 1$ (with $\lambda_{F,S}=2\pi/k_{F,S}$). [Here we have used that $\phi_{+}$ does not vanish for $\varepsilon$ in the range \eqref{range}.]

Similarly, for $\Psi_{2}$ we have the matching conditions
\begin{subequations}
\label{phi1eq}
\begin{align}
&a_{2}\phi_{+}(\varepsilon,p,0)=c_{2}\gamma_{+}+d_{2}\gamma_{-},\label{phi1a}\\
&b_{2}\phi_{-}(-\varepsilon,-p,0)=i\Delta_{0}(c_{2}+d_{2}),\label{phi1b}\\
&a_{2}\phi'_{+}(\varepsilon,p,0)=iq_{+}c_{2}\gamma_{+}+iq_{-}d_{2}\gamma_{-},\label{phi1a2}\\
&b_{2}\phi'_{-}(-\varepsilon,-p,0)=-\Delta_{0}(q_{+}c_{2}+q_{-}d_{2}),\label{phi1b2}
\end{align}
\end{subequations}
with solution
\begin{align}
\frac{c_{2}}{d_{2}}&=-\frac{\gamma_{-}}{\gamma_{+}}\,\frac{q_{-}\phi_{+}(\varepsilon,p,0)+i\phi'_{+}(\varepsilon,p,0)}{q_{+}\phi_{+}(\varepsilon,p,0)+i\phi'_{+}(\varepsilon,p,0)}\nonumber\\
&=-\frac{\gamma_{-}q_{-}}{\gamma_{+}q_{+}}\left[1+{\cal O}(\lambda_{F,S}/l_{m})\right].\label{solutionsatisfies1}
\end{align}

The normalization requirement gives one more equation for each set of coefficients,
\begin{equation}
|a_{n}|^{2}+|b_{n}|^{2}+\frac{q(p)\Delta_{0}^{2}}{m\sqrt{\Delta_{0}^{2}-\varepsilon^{2}}}(|c_{n}|^{2}+|d_{n}|^{2})=1.\label{normeq}
\end{equation}

\subsubsection{Dispersion relation}
\label{dispersionzerothorder}

Since
\begin{equation}
\frac{\gamma_{+}q_{-}}{\gamma_{-}q_{+}}=1+{\cal O}(\varepsilon/\Delta_{0})+{\cal O}(\lambda_{F,S}/\xi_{0}),\label{OOapproximate1}
\end{equation}
we may approximate
\begin{equation}
\frac{c_{1}}{d_{1}}=-1=\frac{c_{2}}{d_{2}}\;\;{\rm for}\;\;\varepsilon\ll\Delta_{0}\;\;{\rm and}\;\;\lambda_{F,S}\ll\xi_{0},l_{m}.\label{OOapproximate2}
\end{equation}
Eqs.\ \eqref{phi2b} and \eqref{phi1b} then give the dispersion relations
\begin{subequations}
\label{disperse12}
\begin{align}
&\varepsilon_{1}(p)={\cal D}^{\rm inv}(p)\equiv\varepsilon_{p},\;\;{\rm for}\;\;\Psi_{1},\label{disperse2}\\
&\varepsilon_{2}(p)=-{\cal D}^{\rm inv}(-p)\equiv -\varepsilon_{-p},\;\;{\rm for}\;\;\Psi_{2},\label{disperse1}
\end{align}
\end{subequations}
with $\varepsilon_{p}$ determined by the equation
\begin{equation}
\phi_{-}(\varepsilon_{p},p,0)=0\Rightarrow {\cal U}\left[\frac{\mu_{-}-\varepsilon_{p}}{\omega_{c}},\sqrt{2}\,pl_{m}\right]=0.\label{phi1varepsequation}
\end{equation}

The dispersion relation of the two modes is plotted in Fig.\ \ref{fig_dispersion}. For small $p$ it is approximately linear, given by
\begin{equation}
\varepsilon_{p}=v_{c}(p-eA_{\rm AR}),\label{epsilonplinear}
\end{equation}
with the definitions
\begin{align}
v_{c}&=1.14\,l_{m}\omega_{c},\;\;
ev_{c}A_{\rm AR}=(\nu_{-}-\tfrac{3}{2})\omega_{c},\nonumber\\
&\Rightarrow A_{\rm AR}=0.88\,(\nu_{-}-\tfrac{3}{2})l_{m}B\equiv -c_{m}l_{m}B,\label{vcpF}\\
\nu_{-}&=\frac{E_{F}+\tfrac{1}{2}g\mu_{\rm B}B}{\hbar\omega_{c}}.\label{numindef}
\end{align}
These results provide the numerical coefficients for $v_{c}$ and $A_{\rm AR}$ in Eqs.\ \eqref{vcvDelta} and \eqref{ABlambdalm}. 

Concerning the coefficient $c_{m}$, we note that, as required by Eq.\ \eqref{range}, the value of $\nu_{-}$ at the Fermi level is in the range $1/2<\nu_{-}<3/2$. The ratio $g\mu_{\rm B}B/\hbar\omega_{c}=gm/2m_{0}$ (with $m_{0}$ the free electron mass) is typically much smaller than unity, so $\nu_{-}\approx 1/2$ will be close to the lower end of this range and $c_{m}\approx 0.88$. The dispersion curves in Fig.\ \ref{fig_dispersion} are plotted for $\nu_{-}=1/2$.

\subsubsection{Eigenstates}
\label{eigenstates}

From the matching conditions we determine the coefficients of the zeroth order eigenstates,
\begin{subequations}
\label{bdeq}
\begin{align}
&\frac{a_{1}}{d_{1}}=\frac{2i\Delta_{0}}{\phi_{+}(-\varepsilon_{p},-p,0)}\equiv Y_{1},\;\;
\frac{a_{2}}{d_{2}}=\frac{-2i\Delta_{0}}{\phi_{+}(-\varepsilon_{-p},p,0)}\equiv Y_{2},\label{bdeqa}\\
&\frac{b_{1}}{d_{1}}=\frac{-2\Delta_{0}q(p)}{\phi'_{-}(\varepsilon_{p},p,0)}\equiv X_{1},\;\;
\frac{b_{2}}{d_{2}}=\frac{2\Delta_{0}q(p)}{\phi'_{-}(\varepsilon_{-p},-p,0)}\equiv X_{2}.\label{bdeqb}
\end{align}
\end{subequations}
It follows that
\begin{equation}
a_{n}/b_{n}={\cal O}(\lambda_{F,S}/l_{m})\ll 1.\label{weightspinup}
\end{equation}
This means that $\Psi_{1}$ and $\Psi_{2}$ in the normal region have most of their weight in the spin-down band, so $\Psi_{1}$ is predominantly an electron state and $\Psi_{2}$ is predominantly a hole state.

The normalization condition \eqref{normeq} simplifies to
\begin{equation}
|b_{n}|^{2}+Y_{0}|d_{n}|^{2}=1,\;\;Y_{0}\equiv 2q(p)\Delta_{0}/m.\label{normsimple}
\end{equation}
Together with Eq.\ \eqref{bdeq} this determines all coefficients (up to an overall phase factor),
\begin{align}
&a_{n}=Y_{n}(X_{n}^{2}+Y_{0})^{-1/2},\;\;b_{n}=X_{n}(X_{n}^{2}+Y_{0})^{-1/2},\nonumber\\
&d_{n}=-c_{n}=(X_{n}^{2}+Y_{0})^{-1/2}.\label{bcdresult}
\end{align}

Because $\phi'_{-}(\varepsilon_{p},p,0)={\cal O}(l_{m}^{-3/2})$, we can estimate
\begin{equation}
\frac{Y_{0}}{X_{n}^{2}}={\cal O}\left(\frac{\xi_{0}\lambda_{F,S}^{2}}{l_{m}^{3}}\right)\ll 1,\label{Y0Xn}
\end{equation}
since we work in the regime where $l_{m}$ is large both compared to $\lambda_{F,S}$ and compared to $\xi_{0}$. We may therefore neglect $Y_{0}$ relative to $X_{n}^{2}$.

\subsection{Inclusion of the Rashba effect}
\label{includeRashba}

We include the Rashba Hamiltonian 
\begin{equation}
\delta H=\begin{pmatrix}
H_{\rm R}&0\\
0&-H_{\rm R}^{\ast}
\end{pmatrix}\label{deltaHdef}
\end{equation}
as a perturbation of the BdG Hamiltonian. To lowest order in this perturbation we need the matrix elements of $\delta H$ in the basis of unperturbed eigenstates $\Psi_{1},\Psi_{2}$. Since $\langle\Psi_{n}\mid\delta H\mid\Psi_{n}\rangle=0$, there is only a single matrix element $\langle\Psi_{2}\mid\delta H\mid\Psi_{1}\rangle=\langle\Psi_{1}\mid\delta H\mid\Psi_{2}\rangle^{\ast}$ to consider. We calculate separately the contributions to this matrix element from the superconducting and normal regions.

\subsubsection{Matrix element in S}
\label{matrixS}

The Rashba Hamiltonian in the superconducting region is
\begin{equation}
\delta H_{S}=\begin{pmatrix}
-i\alpha\tau_{x}\partial_{y}-\alpha p\tau_{y}&0\\
0&-i\alpha\tau_{x}\partial_{y}+\alpha p\tau_{y}
\end{pmatrix}.\label{deltaHS}
\end{equation}
Note that
\begin{equation}
\langle\chi_{\uparrow,\pm}\mid\delta H_{S}\mid\chi_{\downarrow,\pm}\rangle_{S}=0,\label{chiupdown}
\end{equation}
where $\langle\cdots\rangle_{S}$ indicates integration over the superconducting region $y<0$. The matrix element becomes
\begin{align}
&\langle\Psi_{2}\mid \delta H_{S}\mid\Psi_{1}\rangle_{S}=i\alpha\Delta_{0}\sqrt{\Delta_{0}^{2}-\varepsilon^{2}}\nonumber\\
&\quad\times \left[c_{2}^{\ast}d_{1}(1+ip/ q_{-})-c_{1}d_{2}^{\ast}(1+ip/ q_{+})\right].\label{deltaHS2}
\end{align}

With the help of the approximation
\begin{equation}
\frac{c_{2}^{\ast}d_{1}}{c_{1}d_{2}^{\ast}}\approx(\gamma_{-}/\gamma_{+})^{2},\label{approxdeltaHS}
\end{equation}
this gives, for $\varepsilon\ll\Delta_{0}$,
\begin{equation}
\langle\Psi_{2}\mid \delta H_{S}\mid\Psi_{1}\rangle_{S}=\alpha\Delta_{0}^{2}c_{1}d_{2}^{\ast}(p/q_{+}-p/q_{-}).\label{approxdeltaHS2}
\end{equation}

\begin{figure}[tb]
\centerline{\includegraphics[width=0.8\linewidth]{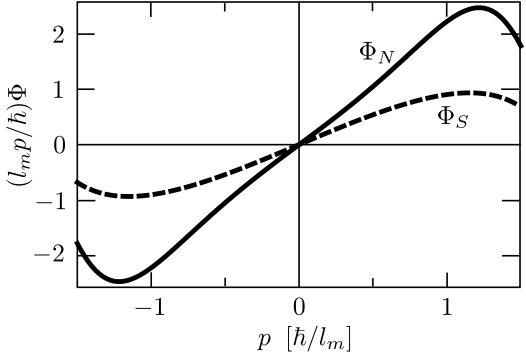}}
\caption{\label{fig_PhiNS}
Plot of the functions $p\Phi_{S}(p)$ and $p\Phi_{N}(p)$, which determine the contribution to the Rashba matrix elements \eqref{Psi12S} and \eqref{deltaHN3} from the superconducting and normal region, respectively. The curves are calculated from Eqs.\ \eqref{PhiSpdef} and \eqref{PhiNdef}, for Fermi energy $E_{F}=\frac{1}{2}\hbar\omega_{c}$. These two functions are of the same order of magnitude, but the contribution to the Rashba matrix element from S has an additional prefactor $(k_{F,S}l_{m})^{-2}$, so it is much smaller than the contribution from N.\\
}
\end{figure}

Since $q_{\pm}=\mp p_{F,S}[1+{\cal O}(\lambda_{F,S}/\xi_{0})+{\cal O}(p/p_{F})^{2}]$, we may further approximate 
\begin{align}
\langle\Psi_{2}\mid \delta H_{S}\mid\Psi_{1}\rangle_{S}&=-2\alpha\Delta_{0}^{2}c_{1}d_{2}^{\ast}\frac{p}{p_{F,S}}
=\frac{2\alpha\Delta_{0}^{2}p}{p_{F,S}|X_{1}X_{2}|}\nonumber\\
&=\frac{\alpha p}{(k_{F,S}l_{m})^{3}}\Phi_{S}(p),\label{Psi12S}
\end{align}
where we have used Eqs.\ \eqref{bcdresult} and \eqref{Y0Xn}, and we have introduced a dimensionless even function of $p$,
\begin{equation}
\Phi_{S}(p)=\tfrac{1}{2}l_{m}^{3}|\phi'_{-}(\varepsilon_{-p},-p,0)\phi'_{-}(\varepsilon_{p},p,0)|.\label{PhiSpdef}
\end{equation}
See Fig.\ \ref{fig_PhiNS} for a plot of $p\Phi_{S}(p)$, which is an approximately linear function of $p$, given for small $p$ by
\begin{equation}
l_{m}p\Phi_{S}(p)=1.13\,l_{m}p+{\cal O}(l_{m}p)^{2}.\label{pPhiSappr}
\end{equation}

\subsubsection{Matrix element in N}
\label{matrixN}

The Rashba Hamiltonian in the normal region is
\begin{widetext}
\begin{equation}
\delta H_{N}=\begin{pmatrix}
-i\alpha\tau_{x}\partial_{y}-\alpha (eBy+p)\tau_{y}&0\\
0&-i\alpha\tau_{x}\partial_{y}-\alpha (eBy-p)\tau_{y}
\end{pmatrix}.\label{deltaHN}
\end{equation}
The matrix element is
\begin{align}
\langle\Psi_{2}\mid \delta H_{N}\mid\Psi_{1}\rangle_{N}={}&i\alpha a_{2}^{\ast}b_{1}\langle\phi_{+}(-\varepsilon_{-p},p,y)\mid -\partial_{y}+ eBy+p\mid\phi_{-}(\varepsilon_{p},p,y)\rangle_{N}\nonumber\\
&+i\alpha b_{2}^{\ast}a_{1}\langle \phi_{-}(\varepsilon_{-p},-p,y)\mid -\partial_{y}-eBy+p\mid\phi_{+}(-\varepsilon_{p},-p,y)\rangle_{N},\label{deltaHN2}
\end{align}
\end{widetext}
with the coefficients given by Eq.\ \eqref{bcdresult},
\begin{equation}
a_{2}^{\ast}b_{1}=-\frac{Y_{2}X_{1}}{|X_{1}X_{2}|},\;\;b_{2}^{\ast}a_{1}=\frac{X_{2}Y_{1}}{|X_{1}X_{2}|}.\label{coeffmatrixN}
\end{equation}

The matrix element can be written in the form
\begin{equation}
\langle\Psi_{2}\mid \delta H_{N}\mid\Psi_{1}\rangle_{N}=\frac{\alpha p}{k_{F,S}l_{m}}\Phi_{N}(p),\label{deltaHN3}
\end{equation}
in terms of a dimensionless even function of $p$,
\begin{subequations}
\label{PhiNdef}
\begin{align}
&\Phi_{N}(p)=\Psi(p)+\Psi(-p),\label{PhiNdefa}\\
&\Psi(p)=\frac{1}{p}\frac{|\phi'_{-}(\varepsilon_{-p},-p,0)\phi'_{-}(\varepsilon_{p},p,0)|}{\phi'_{-}(\varepsilon_{p},p,0)\phi_{+}(-\varepsilon_{-p},p,0)}\nonumber\\
&\times\langle\phi_{+}(-\varepsilon_{-p},p,y)\mid \frac{y}{l_{m}}+pl_{m}-l_{m}\partial_{y}\mid\phi_{-}(\varepsilon_{p},p,y)\rangle_{N}.\label{Psipdef}
\end{align}
\end{subequations}
In Fig.\ \ref{fig_PhiNS} we have also plotted $p\Phi_{N}(p)$. The $p$-dependence is approximately linear, given for small $p$ by
\begin{equation}
l_{m}p\Phi_{N}(p)=c_{N}\,l_{m}p+{\cal O}(l_{m}p)^{2}.\label{pPhiNappr}
\end{equation}
The coefficient $c_{N}$ is a function of $E_{F}/\hbar\omega_{c}$, of order unity. For $E_{F}=\frac{1}{2}\hbar\omega_{c}$ (Fermi level half-way between the splin-split lowest Landau level) one has $c_{N}=1.98$.

\subsection{Andreev-Rashba edge states at an abrupt NS interface}
\label{abruptresults}

From Eqs.\ \eqref{Psi12S} and \eqref{deltaHN3} we find that the matrix element of the Rashba Hamiltonian in the unperturbed basis is 
\begin{equation}
\langle\Psi_{2}\mid\delta H\mid\Psi_{1}\rangle=\frac{\alpha p}{k_{F,S}l_{m}}\bigl[\Phi_{N}(p)+(k_{F,S}l_{m})^{-2}\Phi_{S}(p)\bigr].\label{Psi2Psi1fulla}
\end{equation}
Since both functions $\Phi_{N}(p)$ and $\Phi_{S}(p)$ are of order unity for $p$ of order $1/l_{m}$, the effect of spin-orbit coupling in the superconductor on the Andreev-Rashba edge states is weaker by a factor $1/(k_{F,S}l_{m})^{2}\simeq \hbar\omega_{c}/E_{F,S}$ than the effect of spin-orbit coupling in the normal region. We therefore arrive at the final result for the Rashba matrix element,
\begin{equation}
\langle\Psi_{2}\mid\delta H\mid\Psi_{1}\rangle=\frac{\alpha p}{k_{F,S}l_{m}}\Phi_{N}(p).\label{Psi2Psi1full}
\end{equation}

To first order in the Rashba coefficient $\alpha$, the BdG Hamiltonian in the unperturbed basis is a $2\times 2$ matrix ${\cal H}$ with elements
\begin{equation}
{\cal H}=\begin{pmatrix}
\varepsilon_{p}&(\alpha p/k_{F,S}l_{m})\Phi_{N}(p)\\
(\alpha p/k_{F,S}l_{m})\Phi_{N}(p)&-\varepsilon_{-p}
\end{pmatrix}.\label{HPsi1Psi2}
\end{equation}
The matrix elements have an approximately linear $p$-dependence,
\begin{equation}
{\cal H}\approx\begin{pmatrix}
v_{c}(p-eA_{\rm AR})&v_{\Delta}p\\
v_{\Delta}p&v_{c}(p+eA_{\rm AR})
\end{pmatrix},\label{HPsi1Psi2linear}
\end{equation}
with coefficients $v_{c}$ and $A_{\rm AR}$ given by Eq.\ \eqref{vcpF}. The coefficient $v_{\Delta}$ follows from Eq.\ \eqref{pPhiNappr},
\begin{equation}
v_{\Delta}=c_{N}\frac{\alpha}{k_{F,S}l_{m}}\simeq\frac{v_{c}}{k_{F,S}l_{\rm so}},\label{vDeltaresult}
\end{equation}
in terms of the spin-orbit scattering length $l_{\rm so}=\hbar^{2}/m\alpha$.

The dispersion relation of the Andreev-Rashba edge states, to second order in the Rashba coefficient $\alpha$, is given by
\begin{equation}
\varepsilon_{\pm}=v_{c}p\pm\sqrt{(ev_{c}A_{\rm AR})^{2}+(v_{\Delta}p)^{2}},\label{dispersionpm}
\end{equation}
with the $+$ sign for the electron-like mode $\Psi_{1}$ and the $-$ sign for the hole-like mode $\Psi_{2}$.

\subsection{Andreev-Rashba edge states at a smooth NS interface}
\label{smoothresults}

So far we have taken an abrupt model for the NS interface, with a step function both in the pair potential (from $0$ to $\Delta_{0}$) and in the electrostatic potential (from $0$ to $-V_{0}$). We now turn to the more realistic model of a smooth interface. Since $\Delta_{0}\ll E_{F,S}$ we do not expect the abruptness of the pair potential step to have signficant consequences, so we keep the step function $\Delta(y)=\Delta_{0}\theta(-y)$. 

The situation is different for the electrostatic potential step, which enforces normal reflections at the expense of Andreev reflections. We therefore broaden the step in $V(y)$ over a distance $d$, such that $V(y)=-V_{0}$ for $y\lesssim -d$ and $V(y)=0$ for $y\gtrsim 0$. The abrupt limit corresponds to $d\simeq 1/k_{F,S}$. We now take $d$ larger, but still small compared to $\xi_{0}$.

\subsubsection{Eigenstates in S}
\label{eigenstatesSsmooth}

The eigenstates in N are unaffected by the smoothing for $y<0$. The eigenstates in S are given by $\chi_{s,\pm}w_{\pm}(y)$, with the same spinor $\chi_{s,\pm}$ defined in Eq.\ \eqref{chipmdef} and a spatial profile $w_{\pm}(y)$ determined by
\begin{align}
&-\partial_{y}^{2}w_{\pm}(y)=2m\left[-\mu(p,y)\pm i\sqrt{\Delta_{0}^{2}-\varepsilon^{2}}\right] w_{\pm}(y),\label{wnewdef}\\
&\mu(p,y)=p^{2}/2m+V(y)-E_{F}.\label{mudef}
\end{align}

Since we assume $d\ll\xi_{0}$ we may solve the scattering by the potential step independently of the reflection from the pair potential. The wave vector (in the limit $\varepsilon\rightarrow 0$) changes from $k(p)=\sqrt{2mE_{F}-p^{2}}$ at $y=0$ to $k'(p)=\sqrt{2m(E_{F}+V_{0})-p^{2}}$ at $y=-d$. Plane wave solutions $a_{+}e^{iky}+a_{-}e^{-iky}$ at $y=0$ are related to plane wave solutions $a'_{+}e^{ik'y}+a'_{-}e^{-ik'y}$ at $y=-d$ by a unitary scattering matrix,
\begin{equation}
\begin{pmatrix}
a_{+}\\
a'_{-}
\end{pmatrix}=
\begin{pmatrix}
r&t\\
t'&r'
\end{pmatrix}
\begin{pmatrix}
a_{-}\\
a'_{+}
\end{pmatrix}.\label{smat}
\end{equation}

The solution $w_{\pm}(y)$ corresponds to setting $a'_{\mp}=0$. We thus obtain
\begin{equation}
\frac{w'_{+}(0)}{iw_{+}(0)}=k(p)\frac{r-1}{r+1},\;\;
\frac{w'_{-}(0)}{iw_{-}(0)}=-k(p)\frac{r^{\ast}-1}{r^{\ast}+1},\label{wpmresult}
\end{equation}
with $w'_{\pm}=dw_{\pm}/dy$. The complex conjugation appears as a result of inversion of the scattering matrix, but it can be ignored because the reflection amplitude $r$ is real for $kd\ll 1$.

\subsubsection{Matching at the NS interface}
\label{matchingsmooth}

Matching of the eigenstates in N to those in S at $y=0$ proceeds entirely as in the case of the abrupt interface, with $q_{\pm}$ replaced by the logarithmic derivative $w'_{\pm}(0)/iw_{\pm}(0)$. The result \eqref{bdeq} for the matching coefficients changes simply by the replacement of $q(p)$ by
\begin{equation}
q_{\rm eff}(p)=\frac{w'_{-}(0)}{iw_{-}(0)}=k(p)\frac{1-r}{1+r}.\label{qeffdef}
\end{equation}
The Fermi wave vector $k_{F,S}=q(0)$ is replaced by
\begin{equation}
k_{\rm eff}=q_{\rm eff}(0)=k(0)\frac{1-r}{1+r}.\label{keffdef}
\end{equation}

The reflection amplitude $r$ is related by $r=-\sqrt{1-{\cal T}}$ to the over-barrier transmission probability ${\cal T}$. Since $d\ll\xi_{0}$ and $k\simeq 1/l_{m}\ll 1/\xi_{0}$, we necessarily have $kd\simeq d/l_{m}\ll 1$. One may then expand
\begin{equation}
{\cal T}=c_{\rm barrier}kd+{\cal O}(kd)^{2},\label{calTdef} 
\end{equation}
with $c_{\rm barrier}$ a numerical coefficient of order unity. The wave vector $k_{\rm eff}$ takes the form
\begin{equation}
k_{\rm eff}=\frac{4k(0)}{\cal T}=\frac{4}{c_{\rm barrier}d}.\label{keffresult}
\end{equation}

The value of $c_{\rm barrier}$ depends on the shape of the barrier. As an example, we take the Woods-Saxon step
\begin{equation}
V(y)=-V_{0}\bigl[1+e^{(y+y_{0})/d}\bigr]^{-1},
\end{equation}
with $y_{0}\gg d$ (so that the potential is essentially zero for $y>0$). The transmission probability is\cite{Ahm99}
\begin{align}
{\cal T}&=1-\frac{\sinh^{2}[\pi d(k-k')]}{\sinh^{2}[\pi d(k+k')]}\nonumber\\
&=4\pi kd\,{\rm cotanh}\,(\pi k'd)+{\cal O}(kd)^{2}.\label{TWS}
\end{align}
For a step which is smooth on the scale of $\lambda_{F,S}$ (so $k'd\simeq d/\lambda_{F,S}\gg 1$) we arrive at Eq.\ \eqref{calTdef} with $c_{\rm barrier}=4\pi$, hence $k_{\rm eff}=1/\pi d$. In the opposite regime $k'd\rightarrow 0$ of an abrupt potential step we have ${\cal T}=4k/k'$, hence $k_{\rm eff}=k'(0)$ --- as it should be.

\subsubsection{Dispersion relation}
\label{dispersionsmooth}

The zeroth order dispersion relation, which is independent of $q(p)$, remains unchanged, still given by Eqs.\ \eqref{disperse12} and \eqref{phi1varepsequation}, and also the velocity $v_{c}$ remains given by Eq.\ \eqref{vcpF}.
 
The expression \eqref{Psi2Psi1full} for the Rashba matrix element in the unperturbed basis is changed into
\begin{equation}
\langle\Psi_{2}\mid\delta H\mid\Psi_{1}\rangle=\frac{\alpha p}{k_{\rm eff}l_{m}}\Phi_{N}(p),\label{Psi2Psi1fullsmooth}
\end{equation}
with the same function $\Phi_{N}$ as for the abrupt interface (see Fig.\ \ref{fig_PhiNS}). Once again, the dominant contribution to the matrix element comes from the normal region, with the contribution from the superconducting region smaller by a factor $(k_{\rm eff}l_{m})^{-2}\simeq (d/l_{m})^{2}$.

The BdG Hamiltonian in the unperturbed basis still has the form \eqref{HPsi1Psi2linear}, the only difference appearing in the coefficient $v_{\Delta}$. Instead of Eq.\ \eqref{vDeltaresult} for the abrupt interface it is now given by
\begin{equation}
v_{\Delta}=c_{N}\frac{\alpha}{k_{\rm eff}l_{m}}\simeq\frac{v_{c}d}{l_{\rm so}}.\label{vDeltaresultsmooth}
\end{equation}
This is the result \eqref{vcvDelta} used in the analysis of the spin-triplet Josephson effect.

\end{document}